\documentclass[universe,article,accept,pdftex,moreauthors]{Definitions/mdpi}

\usepackage{amsthm}

\usepackage{wasysym}
\usepackage[none]{hyphenat}
\usepackage{aas_macros}

\firstpage{1} 
\pubvolume{1}
\issuenum{1}
\articlenumber{0}
\pubyear{2025}
\copyrightyear{2025}
\externaleditor{Firstname Lastname}
\datereceived{20 November 2024} 
\daterevised{28 December 2024} 
\dateaccepted{13 January 2025} 
\datepublished{ } 
\hreflink{https://doi.org/} 

\pdfoutput=1

\Title{Elucidating the Dark Energy and Dark Matter Phenomena Within the Scale-Invariant Vacuum (SIV) Paradigm} 

\TitleCitation{Elucidating the Dark Energy and Dark Matter Phenomena Within the Scale-Invariant Vacuum (SIV) Paradigm} 

\Author{Vesselin 
 G. Gueorguiev $^{1,2,3,}$*\orcidV{} and Andre Maeder $^{4}$\orcidA{} }  

\AuthorNames{Vesselin G. Gueorguiev and Andre Maeder}

\AuthorCitation{Gueorguiev, V.G.; Maeder, A.}

\address{%
$^{1}$ \quad Institute for Advanced Physical Studies, 1784 Sofia, Bulgaria\\
$^{2}$ \quad National Coalition of Independent Scholars,  Battleboro, VT 05301, 
 USA \\ 
$^{3}$ \quad Ronin Institute for Independent Scholarship, Montclair, NJ 07043, 
 USA\\ 
$^{4}$ \quad Geneva Observatory, University of Geneva, Geneva, CH-1290 Sauverny, 
 Switzerland; 
andre.maeder@unige.ch 
}

\corres{Correspondence: vesselin@mailaps.org} 

\abstract{The enigmatic phenomenon of dark energy (DE) is 
the elusive entity driving the accelerated expansion of our Universe. 
A plausible candidate for DE is the non-zero Einstein Cosmological Constant $\Lambda_{E}$ 
manifested as a constant energy density of the vacuum, yet it seemingly defies gravitational effects. 
In this work, we interpret the non-zero $\Lambda_{E}$ through the lens of scale-invariant cosmology. 
We revisit the conformal scale factor $\lambda$ and its defining equations within 
the Scale-Invariant Vacuum (SIV) paradigm. 
Furthermore, we address the profound problem of the missing mass across galactic and extragalactic scales by deriving an MOND-like relation, 
$g \sim \sqrt{a_0\,g_N}$, within the SIV context. Remarkably, the values obtained for $\Lambda_{E}$ and the MOND fundamental acceleration, 
$a_0$, align with observed magnitudes, specifically, \mbox{$a_0 \approx 10^{-10} \, \mathrm{m} \, \mathrm{s}^{-2}$ and 
$\Lambda_{E} \approx 1.8 \times 10^{-52} \, \mathrm{m}^{-2}$.}
Moreover, we propose a novel early dark energy term, $\tilde{T}_{\mu\nu} \sim \kappa H$, 
within the SIV paradigm, which holds potential relevance for addressing the Hubble tension. 
}

\begin{document}
\tableofcontents 

\section{Introduction}

Modern physics faces a tantalizing situation wherein the two 
theories describing most phenomena, Quantum Field Theory (QFT) and 
General Relativity (GR), have been successfully tested to high precision 
on Earth and via solar system observations. 
However, the models of phenomena on galactic, inter-galactic, 
and cosmic scales have suggested that normal matter 
barely accounts for $\approx$5\% of the energy 
content of the Universe. Meanwhile, $\approx$70\% of the 
energy content is related to the expansion of the Universe due to dark energy (DE), and the other component is $\approx$25\%
due to dark matter (DM) \cite{CosmolParameters-Planck'15results}.

Following the conventional wisdom,
there is no shortage of proposals as to what DE and DM could be:
either possible new fields, new particles, 
or modifications of the Einstein GR (EGR) \cite{DE'08,DE'11,ExtendedGravity'11}. 
While one is often very successful in continuing an ongoing trajectory,
the usual explanation of DE and DM has faced a detection deficit for over 40 years
while accumulating many tensions as a working paradigm \cite{Kroupa:2023ubo}.

Here, we would like to provide a possible solution to the 
DE and DM puzzles via a symmetry extension of the EGR.
This extension was already proposed by Weyl in 1918~\cite{Weyl-Geometry-Weyl'18-ed6,Weyl-Geometry-Weyl'22-English},
but was rejected, for good reason, by Einstein \cite{Weyl-Geometry-Objection-Einstein'18}.
The initial concern was resolved within the Weyl integrable geometry (WIG)
reformulation \cite{Eddington'23,Weyl-Geometry-Dirac'73}. 
The idea of scale invariance was then advocated by \citet{Weyl-Geometry-Dirac'73}, and 
it was later the cornerstone of the scale-covariant cosmology of \citet{Scale-covariance-Canuto'77}.
The initial works \cite{Weyl-Geometry-Dirac'73, Scale-covariance-Canuto'77} 
used the Large Numbers Hypothesis of \citet{Dirac-LargeNumbers'74},
which was faced with skepticism. However, in 2016, the idea of 
a Scale-Invariant Vacuum (SIV) was proposed 
 in three papers posted to the arXiv preprint server in 2016
and published in 2017 by \citet{Maeder'17a}. 
Since then, the idea has been stress tested on various phenomena.
For a short, recent overview, {see} 
 \cite{SIV-PartII'24}. 
The foundations of the framework have been revisited \cite{SIVandDS'20,SIV-PartI'23},
and the potential link between the SIV and DM and DE was stated in 2020 by \citet{SIV-Origin-of-DMandDE'20}.
Here, we provide our current understanding of the phenomena 
of dark energy via the Einstein Cosmological Constant 
and dark matter via the MOND-like acceleration relation within the SIV paradigm, 
along with the numerical values of the relevant expressions.

\section{Framework for Scale-Invariant Cosmology}

\textls[-15]{As was already pointed out, 
the Weyl geometry as the foundation of a scale-invariant cosmology 
was discussed by \citet{Weyl-Geometry-Dirac'73} and \citet{Scale-covariance-Canuto'77},
and possible astronomical applications 
have been considered by \citet{WIG'78} and \citet{ScaleInv_and_AstroMotion'79}.
More modern but very abstract mathematical formulations 
have been reviewed~\cite{WeylGeom'18}; 
however, nature-based considerations of phenomena with 
potential observational validations are rare 
\cite{ScaleInv_and_AstroMotion'79,ScaleInvGravity-and-BH-ringdown'20,SIV-PartII'24}, 
and even fewer have had a particle physics focus~\cite{Conformal_Spaces'73}.}

\textls[-23]{The scale-covariant cosmology equations were first introduced 
in 1977 by \mbox{\citet{Scale-covariance-Canuto'77}}} in the following form:

\begin{eqnarray}
\frac{8\,\pi G\varrho}{3}=\frac{k}{a^{2}}+\frac{\dot{a}^{2}}{a^{2}}+2\,\frac{\dot{\lambda}\,\dot{a}}{\lambda\,a}
+\frac{\dot{\lambda}^{2}}{\lambda^{2}}-\frac{\Lambda_{\mathrm{E}}\lambda^{2}}{3}\,,\label{E1p}\\
-8\,\pi Gp=\frac{k}{a^{2}}+2\frac{\ddot{a}}{a}+2\frac{\ddot{\lambda}}{\lambda}+\frac{\dot{a}^{2}}{a^{2}}
+4\frac{\dot{a}\,\dot{\lambda}}{a\,\lambda}-\frac{\dot{\lambda^{2}}}{\lambda^{2}}-\Lambda_{\mathrm{E}}\,\lambda^{2}\,.\label{E2p}
\end{eqnarray}

As a scale-invariant cosmology, any $\lambda$ could {be used}. Thus, one has to make a ``gauge'' choice for $\lambda$ 
to proceed with comparison to 
observations\endnote{As it stands, one can fit observations and deduce the model parameters, 
but the choice of $\lambda$ has to make sense from a physics viewpoint.}.  
Some possible choices 
have already been discussed by \citet{Scale-covariance-Canuto'77}
based on the Large Numbers Hypothesis by \citet{Dirac-LargeNumbers'74}.
Subsequent studies, e.g., \cite{ScaleInv_and_AstroMotion'79}, 
noticed that, for the vacuum solutions of the GR equations,
the conformal equivalence of de Sitter space 
to Minkowski space can be achieved explicitly by fixing $\lambda$ to satisfy
\begin{equation}
3\lambda^{-2}/(c^2t^2\Lambda_E)=1.\label{lambda}
\end{equation}

This property of the vacuum, that {\it {the empty space is scale invariant}}, 
was further formalized in~\cite{Maeder'17a}, and, 
in subsequent works, was emphasized as the Scale-Invariant Vacuum hypothesis for 
imposing the SIV gauging condition that fixes $\lambda$. Therefore, 
if one is to choose an $\lambda$ that does not depend on the matter behavior explicitly, 
and by setting $\Lambda=\Lambda_E\lambda^2$, 
then one can obtain the following set of relationships given first in 
\cite{Maeder'17a} and  further re-derived from an action in \cite{SIV-PartI'23},
while the observational consequences have been summarized in \cite{SIV-PartII'24}:

\begin{eqnarray}
\ 3\,\frac{\dot{\lambda}^{2}}{\lambda^{2}}\,=\Lambda\,,\quad\mathrm{and}\quad2\frac{\ddot{\lambda}}{\lambda}
-\frac{\dot{\lambda}^{2}}{\lambda^{2}}\,=\Lambda\,,\label{SIV1}\\
\mathrm{or}\quad\frac{\ddot{\lambda}}{\lambda}\,=\,2\,\frac{\dot{\lambda}^{2}}{\lambda^{2}}\,,
\quad\mathrm{and}\quad\frac{\ddot{\lambda}}{\lambda}-\frac{\dot{\lambda}^{2}}{\lambda^{2}}\,=\frac{\Lambda}{3}\,.\label{SIV2}
\end{eqnarray}

Upon the use of the SIV choice \eqref{SIV1} first introduced in 2017 by \citet{Maeder'17a} 
or its equivalent form \eqref{SIV2}, one observes that 
{\it {the cosmological constant disappears}} from \eqref{E1p} and \eqref{E2p}. 
In doing so, one recovers the scale invariance of the vacuum for flat cosmology ($k=0$), 
which is broken only by the presence of source fields 
characterized by the energy density of matter and its~pressure:

\begin{eqnarray}
\frac{8\,\pi G\varrho}{3}=\frac{k}{a^{2}}+\frac{\dot{a}^{2}}{a^{2}}+2\,\frac{\dot{a}\dot{\lambda}}{a\lambda}\,,\label{E1}\\
-8\,\pi Gp=\frac{k}{a^{2}}+2\frac{\ddot{a}}{a}+\frac{\dot{a^{2}}}{a^{2}}+4\frac{\dot{a}\dot{\lambda}}{a\lambda}\,.\label{E2}
\end{eqnarray}

Making sense of this choice of $\lambda$, called the Scale-Invariant Vacuum (SIV) choice, 
and linking it to the observed phenomena in nature are the purposes of this paper.
For this purpose, we go back to the origin of the equations and
highlight the key properties of the Ricci tensor and scalar 
related to understanding our viewpoint.
After a Weyl transformation \cite{WIG'78,ScaleInv_and_AstroMotion'79}, 
one has the following Ricci tensor and Ricci scalar expressions:
\begin{equation}
g_{\mu\nu}\rightarrow\lambda^{2}g_{\mu\nu}
\Rightarrow R_{\mu\nu}\rightarrow R_{\mu\nu}+K_{\mu\nu},
\label{eq:metric-conf-transf}
\end{equation}
where $K_{\mu\nu}$ is given by
\begin{equation}
K_{\mu\nu}=g_{\mu\nu}\kappa^{\rho}\kappa_{\rho}+2\kappa_{\mu}\kappa_{\nu}
+\kappa_{\mu;\nu}+\kappa_{\nu;\mu}-2g_{\mu\nu}\kappa_{;\rho}^{\rho}.
\label{eq:K-tensor}
\end{equation}

Such expressions were first discussed in Eq. (89.2) by 
\citet{Eddington'23} and later 
by \citet{Weyl-Geometry-Dirac'73} and \citet{Conformal_Spaces'73} as well.
Upon contracting \eqref{eq:metric-conf-transf}, one sees that
the Ricci scalar becomes $R\rightarrow (R+K)/\lambda^2$, where
\begin{equation}
K=6\kappa^{\rho}\kappa_{\rho}-6\kappa_{;\rho}^{\rho}.
\label{eq:K}
\end{equation}

Here, $\kappa_{\mu}=-\partial_{\mu}\ln\lambda$ is the WIG connexion vector. 

\section{The Various Faces of the Cosmological Term $\Lambda$}

\subsection{The Einstein Cosmological Constant $\Lambda_{E}$}\label{Constant ECC}

The conventional Einstein equation with cosmological constant $\Lambda_{E}$
within general relativity is as follows:
\begin{equation}
R_{\mu\nu}-\frac{1}{2}Rg_{\mu\nu}+\Lambda_{E}g_{\mu\nu}=\varkappa T_{\mu\nu}.
\label{eq: Einstein}
\end{equation}

This formulation utilizes a metric-compatible connection 
where the first two terms form the Einstein tensor 
with zero-divergence 
while on the right-hand side. When on the left-hand side, one has the stress--energy tensor satisfying similar 
zero divergence, leading to the relevant covariant conservation laws. 
All of this is consistent as long as $\varkappa=8\pi G/c^{4}$ and the cosmological constant $\Lambda_{E}$
are constants; otherwise, one would have $\Lambda_{E,\mu}=\varkappa_{,\nu}T_{\mu}^{\nu}$.
Thus, the constancy of $G$ and $c$ imply the constancy of the Einstein Cosmological Constant $\Lambda_{E}$.

\subsection{Cosmological Constant or Dark Energy}

One usually expects that an appropriate averaging over the matter
distribution would result in a stress--energy tensor $T_{\mu\nu}=\rho g_{\mu\nu}+O(\delta T)$
where the energy density $\rho$ will be related to 
the zero-point/vacuum energy of the matter fields. 
Since, in a co-moving frame,
for an observer at infinity, 
the metric is expected to be Minkowski-like, one has 
$T_{00}\varpropto\rho$ and $T_{ii}\varpropto p$; therefore, 
one naturally considers $\rho$  
as a dark energy contribution to the stress--energy tensor with
$p=-\rho$ due to $\eta_{00}=-\eta_{ii}$. Thus, we can move the term
$\Lambda_{E}\,g_{\mu\nu}$ from the LHS to the RHS of (\ref{eq: Einstein})
and consider $\Lambda_{E}$ to be related to the zero-point/vacuum
energy of the matter fields. Unfortunately, this leads to the cosmological
constant problem manifested in the enormous discrepancy between the estimated
value, based on QFT arguments, 
and the observed/measured actual value 
\cite{CCP-Weinberg'89,CCP-Carroll'92,CosmolParameres-WMAP'13,CosmolParameters-Planck'15results}.
Another issue comes from the parallels between Newtonian Gravity, where 
a homogeneous and isotropic mass distribution has no gravitational effect,
and GR, where energy is on the RHS of \eqref{eq: Einstein}
and therefore has an influence on the metric.
Thus, the question of ``Should or shouldn't the vacuum gravitate if it has a non-zero energy density?'' comes to mind.

\subsection{Connecting the Dots Within the SIV Paradigm}

The observed value for $\Lambda_{E}$ is well within the order of
magnitude estimate based on the relevant parameters for such a system \cite{CCP'20},
that is, using the values of $c,G,R_{H}$ where the Hubble radius
is $R_{H}=c/H_{0}$ with $H_{0}$ being the Hubble constant. 
However, if one considers a non-zero 
positive constant energy density for a homogeneous
and isotropic universe, then one concludes that, at a sufficiently
large distance $R_{S}$, such a universe should possess a black hole event horizon. 
For example, if the constant energy density is due to the zero-point energy of the 
familiar matter and radiation fields, then one expects $\rho_0=const \gtrsim 0$.
Now, consider a ball of radius $r$. Such a constant value results in 
an effective mass $M=\frac{4\pi}{3}r^3\rho_0$, 
in geometric units ($c=1, G=1$), and its Schwarzschild radius will be $R_S=2M$.
If the mass distribution is within a radius $r<R_S$, then one has a black hole.
No matter how small the positive constant energy density $\rho_0 \gtrsim 0$ is,
there is always a sufficiently big ball of radius $r_b$,
so that beyond $r>r_b$, one has a black hole with an event horizon at 
$r_b=\frac{8\pi}{3}r_b^3\rho_0\Rightarrow r_b=\sqrt{\frac{3}{8\pi\rho_0}}$.
Of course, this situation does not apply if the constant energy density is negative.
Therefore, one can inevitably conclude that we are inside a black hole, 
which may be the case as argued in \cite{BHU'72, BHU'22}.
However, if it is so, then one would expect a contracting flow of matter towards the 
center of such a black hole, located somewhere in the past, 
where there should be the biggest concentration of matter;
however, we observe an expanding Hubble flow towards the future event horizon,
which would indicate that the constant energy density should be negative. 
Such a negative energy density will result in negative effective mass,
which is at odds with positive probabilities in quantum mechanics 
\cite{RepInv-and-KeyProps-of-Phys-Sys'21}.

An alternative viewpoint is to consider the observed age of the Universe
$\tau_{0}\approx1/H_{0}$ instead. Then again, $c,G,\tau_{0}$ will give
us the ballpark estimate for $\Lambda_{E}$ but with a different
understanding when viewed within the SIV paradigm. For this purpose,
we shall consider the Weyl transformation \eqref{eq:metric-conf-transf}. 
That is, the metric in the Einstein GR (EGR) frame $g'_{\mu\nu}$ will be related
to a metric $g_{\mu\nu}$ within a Weyl integrable geometry (WIG)
via the factor $\lambda$, that is, $g'_{\mu\nu}=\lambda^{2}g_{\mu\nu}$. From now on, we will denote the EGR frame quantities with primes and
no primes for the more general WIG quantities. 
Upon utilizing \eqref{eq:metric-conf-transf}, 
\eqref{eq:K-tensor}, and \eqref{eq:K} within \eqref{eq: Einstein},
such a Weyl transformation expresses \eqref{eq: Einstein} 
into the more general WIG framework and the Einstein equation becomes
\begin{equation}
R_{\mu\nu}+K_{\mu\nu}-\frac{1}{2}\lambda^{-2}(R+K)\lambda^{2}g_{\mu\nu}
+\Lambda_{E}\lambda^{2}g_{\mu\nu}=\varkappa\,T_{\mu\nu}.
\label{eq: Einstein-new}
\end{equation}

The above Equation \eqref{eq: Einstein-new} should be viewed as a rewriting of 
the Einstein GR equations with a cosmological constant \eqref{eq: Einstein} for the
metric $g'_{\mu\nu}$ into extended equations within the WIG framework 
where now the objects on the LHS and RHS of \eqref{eq: Einstein-new}
depend on $\lambda$ and $g_{\mu\nu}$. If $\lambda$ is chosen to be a
constant, then \eqref{eq: Einstein-new} reverts to \eqref{eq: Einstein} 
because the connexion vector $\kappa_{\mu}$ becomes zero and 
so is $K_{\mu\nu}$, as seen from \eqref{eq:K}, 
but for an appropriate new choice of units that is reflected in the appropriate rescaling of $\Lambda_{E}$,
one will have the extended equation \eqref{eq: Einstein-new}. 
In what follows, we are interested in a non-trivial choice for $\lambda$ that will depend on time only. 
To specify the functional form of this non-trivial $\lambda$, 
one can make the following considerations: 
To understand the forthcoming expressions for $\lambda$,
the new Equation \eqref{eq: Einstein-new} 
can now be split into two equations, one containing 
$\Lambda=\lambda^{2}\Lambda_{E}$ along with $\lambda$ and 
its derivatives (via $k_{\mu}$), 
and another equation that does not have the $\Lambda$ term:
\begin{eqnarray}
R_{\mu\nu}-\frac{1}{2}Rg_{\mu\nu}=\varkappa T_{\mu\nu} -\tilde{T}_{\mu\nu},
\label{eq: No Lambda}\\
K_{\mu\nu}-\frac{1}{2}Kg_{\mu\nu}+\Lambda g_{\mu\nu}=\tilde{T}_{\mu\nu}.
\label{eq: SIV1}
\end{eqnarray}

To evaluate $\tilde{T}_{\mu\nu}$, we look at the LHS of \eqref{eq: SIV1} and use \eqref{eq:K-tensor} and \eqref{eq:K}:
$\tilde{T}_{\mu\nu}=2\kappa_{\mu}\kappa_{\nu}+\kappa_{\mu;\nu}+\kappa_{\nu;\mu}+
(\Lambda-2\kappa^{\rho}\kappa_{\rho}+\kappa_{;\rho}^{\rho})g_{\mu\nu}$.
Therefore, imposing $\kappa_{\mu;\nu}=\kappa_{\nu;\mu}=-\kappa_{\mu}\kappa_{\nu}$
along with $\Lambda=3\kappa_{\mu}\kappa^{\mu}$ will guarantee $\tilde{T}_{\mu\nu}=0$.
By looking at $\kappa_{\mu;\nu}=\kappa_{\mu,\nu}+\Gamma^\rho_{\mu\nu}\kappa_{\rho}$,
the first condition is readily satisfied within the WIG ($\kappa_{\mu;\nu}=\kappa_{\nu;\mu}$) of
\citet{Scale-covariance-Canuto'77},
while the second condition implies the relationship 
$\kappa_{\mu,\nu}+\Gamma^\rho_{\mu\nu}\kappa_{\rho}+\kappa_{\mu}\kappa_{\nu}=0$.
For the SIV ``gauge'' choice $\lambda\propto 1/t$, the only non-zero component is $\kappa_0=1/t$, 
which will require $\Gamma^0_{\mu\nu}=0$. In particular, $\Gamma^0_{00}=0$ implies a time-independent $g_{00}$
for constructing the metric-compatible covariant derivative. 
Thus, for a general metric, the LHS of \eqref{eq: SIV1} may result in non-zero 
$\tilde{T}_{\mu\nu}=\Gamma^0_{\mu\nu}\kappa_0$ within the SIV paradigm. 
Therefore, \eqref{eq: SIV1} defines $\tilde{T}_{\mu\nu}$ once the choice of $\lambda$ is made.
In this respect, $\tilde{T}_{\mu\nu}$ can bring early dark energy effects into \eqref{eq: No Lambda}
due to the time dependence of $\kappa_0$. 
The contemporary view on the resolution of the Hubble tension is the possibility for early dark energy \cite{EDEandHT'23},
which could be supplied by $\tilde{T}_{\mu\nu}$.
For the case of SIV theory, the appearance of a non-zero $\tilde{T}_{\mu\nu}$ 
and its dependence on $\Gamma^0_{\mu \nu}$ and $\kappa$ result in the coupling of the Hubble parameter $H=\dot{a}/{a}$
to $\kappa=-\dot{\lambda}/{\lambda}$, as seen by the last terms in \eqref{E1} and \eqref{E2}.
This gives an explicit new model for early dark energy based on $\tilde{T}_{\mu\nu}\sim\kappa\,H$.
From what follows, this split of \eqref{eq: Einstein-new} is viewed 
as a foresight\endnote{The need to fix $\lambda$ has been anticipated by \citet{Weyl-Geometry-Dirac'73,Scale-covariance-Canuto'77},
but they used the Large Numbers Hypothesis \citet{Dirac-LargeNumbers'74}. 
Here, we present the SIV approach, which seems to be relevant for understanding the cosmological constant and the dark matter phenomena. 
Another ``gauge'' fixing is the $\lambda$ constant that is the EGR frame.
There could be other ``gauge'' choices within the WIG that will correspond to specific WIG frameworks. 
The significance of these frameworks is something to be understood in the future.
In particular, the more correct expression for $\Lambda$ in \eqref{eq: Lambda SIV} contains a linear term in $\kappa$,
of the form $\Gamma^\rho_{\rho 0}\kappa_0$,
that comes from $\kappa^\rho_{;\rho}$. 
This term, along with other terms that result in an overall non-zero value for the LHS of \eqref{eq: SIV1}, 
can be part of the stress--energy tensor $\tilde{T}_{\mu\nu}$ determining $g_{\mu\nu}$ via \eqref{eq: No Lambda}.
These extra terms to $T_{\mu\nu}$ could be viewed as dark energy that are not directly related to the cosmological constant.
For example, another metric-specific term is $\varGamma_{0i}^{0}\kappa_0$, which is not balanced in general when considering \eqref{eq:K-tensor}, \eqref{eq:K}, and \eqref{eq: SIV1}. 
Remarkably, all the terms with an explicit $g_{\mu\nu}$ will cancel out upon using 
the more general expression $\kappa^\rho_{\rho;}=-\kappa^\rho\kappa_{\rho}$ instead of 
$\dot{\kappa}=-\kappa^2$, as given in \eqref{eq: Lambda SIV},
but $\dot{\kappa}=-\kappa^2$ is important since it guarantees a constant value for $\Lambda/\lambda^2$
and therefore constancy of $\Lambda_E$.
In this respect, the unique choice for $\lambda$ that follows from \eqref{eq: Lambda SIV},
which is equivalent to \eqref{SIV1} and \eqref{SIV2}, is an equivalent definition of the main SIV equations within a special co-moving frame. 
Furthermore, the SIV theory associated with the unique ``gauge'' choice 
defined by Equation \eqref{SIV1} and/or the equivalent set \eqref{SIV2}
is also supported by the unique scale-invariant action principle discussed recently in \cite{SIV-PartI'23}.
}.
When investigating how to choose the  ``gauge'' factor $\lambda$, the second equation \eqref{eq: SIV1} defines $\tilde{T}_{\mu\nu}$ once the choice of $\lambda$ is made, and one is left only with the first equation \eqref{eq: No Lambda}.

Now, consider $\kappa^\rho_{;\rho}=-\kappa^\rho\kappa_\rho$ along with 
$\Lambda=2\kappa^{\rho}\kappa_{\rho}+\kappa_{;\rho}^{\rho}=3\kappa^\rho\kappa_\rho$,
which implies $\Lambda=K/4$ and $\tilde{T}=0$.
In the special co-moving frame where the time-covariant derivative is given by the partial derivative,
with $\lambda$ dependent only on time and $\kappa=\kappa_{0}=-\dot{\lambda}/\lambda$, 
one obtains a key SIV equation:
\begin{equation}
\Lambda=\frac{3}{2}\left(\kappa^{2}-\dot{\kappa}\right)\Leftrightarrow\Lambda=3\kappa^{2}
=3\left(\frac{\dot{\lambda}}{\lambda}\right)^{2}:\mathrm{iff}\,\:\dot{\kappa}=-\kappa^{2}
\label{eq: Lambda SIV}
\end{equation}

The above equations are those given by the first expressions in \eqref{SIV1} and \eqref{SIV2}.
By taking the time derivative of $\kappa/\lambda$, we see that it will
be equal to $(\dot{\kappa}+\kappa^{2})/\lambda$, which will vanish if
$\dot{\kappa}=-\kappa^{2}$; thus, $\Lambda/\lambda^{2}$ will
be a constant\endnote{The SIV equations for $\lambda$ have been 
redirived from an action principle \cite{SIV-PartI'23},
but were first introduced and studied, since 2017, by \citet{Maeder'17a}, 
within the scale-invariant cosmology by
\citet{Weyl-Geometry-Dirac'73,Scale-covariance-Canuto'77}.
Thus, the property of $\dot{\kappa}=-\kappa^{2}$  
has been noticed before and in particular the result within the SIV that 
$\dot{\lambda^{2}}/\lambda^{4}$ is constant. 
Here, we turn this observations into a reasonable choice 
for determining the functional form of $\lambda$ 
that results in $\Lambda_{E}$ being a constant according to the SIV.}.
Therefore, we will denote this constant also by $\Lambda_{E}$ 
since these two constants will coincide eventually.
Thus, for the case $\dot{\kappa}=-\kappa^{2}$, the
solution for $\lambda(t)$ is very simple and one can use the constant
$\Lambda_{E}=\Lambda/\lambda^{2}=3\dot{\lambda^{2}}/\lambda^{4}$
to characterize~it: 
\begin{equation}
\varepsilon(t_{0}-t)\sqrt{\Lambda_{E}/3}=1/\lambda-1/\lambda_{0}.
\label{eq: Constant CC}
\end{equation}

That is, $\kappa^{2}=\dot{\lambda}^{2}/\lambda^{2}=\Lambda/3=\left(\Lambda_{E}/3\right)\lambda^{2}$,
along with 
\begin{equation}
\lambda=\lambda_{0}/\left(1+\lambda_{0}\varepsilon(t_{0}-t)\sqrt{\Lambda_{E}/3}\right),
\label{eq: lambda}
\end{equation}
and $\kappa=-\dot{\lambda}/\lambda=-\varepsilon\lambda\sqrt{\Lambda_{E}/3}$
are key SIV expressions; 
therefore, $\dot{\kappa}\sim\dot{\lambda}=\varepsilon\lambda^{2}\sqrt{\Lambda_{E}/3}$
implies $\dot{\kappa}=-\varepsilon^{2}\left(\Lambda_{E}/3\right)\lambda^{2}=-\kappa^{2}$
as required. Here, $\varepsilon=\pm1$, and therefore $\varepsilon^{2}=1.$
By setting $\lambda_{0}\sqrt{\Lambda_{E}/3}=1/t_{0}$, we have $\lambda=\lambda_{0}t_{0}/t$
for $\varepsilon=-1$; 
thus, we have recovered \eqref{lambda}, and therefore 
$\kappa=-\dot{\lambda}/\lambda=1/t$
with $t\in[t_\text{in},t_{0}]$, where $t_\text{in}$ is the moment of the Big Bang
when $a(t_\text{in})=0$. 

It is often convenient to choose $\lambda_{0}=1$
along with SIV time units such that $t_{0}=1$.
Thus, one is led to the same expression of the scale factor $\lambda$ 
as obtained by the fundamental SIV hypothesis, 
according to which the macroscopic empty space is scale invariant, 
homogeneous, and isotropic \citep{Maeder'17a, SIV-PartI'23}.

\subsection{Interpretation of the Cosmological Constant Within the SIV Framework}

The presence of a non-zero cosmological term $\Lambda_{E}$ leads to some severe problems:
(1)~a mismatch of the observed value with the zero-point energy estimates based on QFT,
and (2) the puzzling conclusion that we may be inside a black hole. 
The Quantum Field Theory (QFT) predicts an enormous value of vacuum energy 
when viewed as the zero-point energy of the matter fields (i.e., $c^7/\hbar/G^2 \sim 10^{114}$ erg/cm$^2$), 
while anthropic considerations à la Weinberg and even a simple dimensional estimate 
using the relevant physical constants (i.e.,  $H_0^2c^2/G\sim10^{-8}$ erg/cm$^2$) 
seem to arrive at the correct order of magnitude for the vacuum energy related to the cosmological constant. 
Thus, one can conclude that quantum effects involving Planck’s constant $\hbar$ have nothing to do with the observed $\Lambda_{E}$. 
Therefore, quantum vacuum fluctuations are just that: fluctuations whose mean value is zero at large cosmic scales. 
Within the SIV, this is reflected in removing the $\Lambda_{E}$ from the Freedman equations, as seen in \eqref{eq: No Lambda},
in favor of an early dark energy term defined by \eqref{eq: SIV1} that involves the conformal factor $\lambda$. 
It can be interpreted as a choice of parameterization that brings the GR equations 
into the true co-moving frame with no cosmological constant $\Lambda_{E}$ as extra energy density. 
It is similar to what happens when 
identifying the co-moving frame such that the kinetic energy of a system is zero and therefore
there is no relative special motion.
However, we do leave in 4D spacetime, which brings up the question of relative time parameterizations;
that is, what if the coordinate time of the observer is different from the proper time of the system under study?
It seems the relative time parametrization controlled by $\lambda$ also controls the amount of extra energy that there could be.

Another way to understand the situation is to recognize that the positive cosmological constant $\Lambda_{E}$
on the LHS of \eqref{eq: Einstein} indicates extra energy density as part of the RHS \eqref{eq: Einstein}.
The presence of $\Lambda_{E}$ explicitly breaks the global rescaling symmetry along with the 
$\rho$, $p$, and $k/a^2$ terms in the Freedman equations. 
The breaking is still there even for the macroscopic vacuum, characterized by $\rho=p=k=0$, if $\Lambda_{E}$ is non-zero. 
This can be viewed as a manifestation of {\it {unproper}} 
time parametrization, since, for proper time parametrization, one expects zero energy density instead. 
To correct the time parametrization, one can apply global conformal transformation $\lambda(t)$
instead of the commonly discussed local conformal gauge $\lambda(x)$. 
The use of $\lambda(x)$ would imply the presence of a physical field whose 
excitations should manifest as particles, which is not permissible \cite{Conformal_Spaces'73}.
Thus, the idea of using $\lambda(t)$ is well justified in order to preserve isotropy and homogeneity of space.
It is aligned with the idea about the role of time parametrization. 
Therefore, the existence of $\lambda(t)$ as defined by \eqref{eq: lambda} 
removes $\Lambda_{E}$ from the Freedman equations and 
results in \eqref{E1} and \eqref{E2}, which are clearly scale invariant when $\rho=p=k=0$.
This demonstrates the relationship between the scale-breaking term 
$\Lambda_{E}$ and its relation to the symmetry-restoring WIG frame defined by $\lambda(t)$ given by \eqref{eq: lambda}.

Therefore, the ``gauge'' symmetry of the SIV theory 
is not like the usual local gauge symmetry, which we are familiar with from particle physics.
As such, one can circumvent the earlier mentioned problems  
by showing that $\Lambda_{E}$ is an actual constant within the SIV.
As mentioned earlier, the Einstein Cosmological Constant $\Lambda_{E}$ 
must be a constant if one views the Newton Gravitational Constant $G$ as a true constant
(see Section \ref{Constant ECC}).
This can be used to construct a Weyl transformation that
removes the cosmological term $\Lambda$. It implies that the extra
energy density due to the cosmological constant can be viewed as an observer
effect, just as the kinetic energy of a system depends on the relative
motion of the two systems; however, in the case of the cosmological
constant, this seems to be about the difference in time parameterizations.
That is, the metric $g_{\mu\nu}$ provides $\Lambda$-free EGR equations,
as seen in (\ref{eq: No Lambda}), while the presence of a non-zero constant
$\Lambda_{E}$ term is due to the choice of the EGR metric tensor
$g'_{\mu\nu}=\lambda^{2}g_{\mu\nu}$, where the factor $\lambda=t_{0}/t$
is defined via $\sqrt{\Lambda_{E}/3}=1/t_{0}$ \eqref{eq: lambda},
in agreement with the early observation by \cite{ScaleInv_and_AstroMotion'79, Maeder'17a} 
that one can obtain the Minkowski line element based on \eqref{lambda}. 
Thus, this relates the value of the cosmological constant $\Lambda_{E}$ to the age of the Universe $t_{0}$, 
which is consistent with the simple dimensional estimates \cite{CCP'20}.  
In the usual SI units where the age of the Universe is $\tau_{0}=13.8$ billion years
and the speed of light is $c=3\times10^8$ m/s, one obtains
\begin{equation}
\Lambda_{E}=3/(c\tau_0)^2\approx 1.8\times 10^{-52}\;\text{m}^{-2},
\label{LambdaE}
\end{equation}
which is reasonably close to the measured value
\cite{CCP'20,CosmolParameters-Planck'15results}.

In the considerations above, we have settled that the time dependence
of $\lambda$ only is justifiable based on assumptions of 
homogeneous and isotropic space at cosmological scales.
This is a cornerstone of the SIV paradigm \cite{Maeder'17a,SIV-PartI'23}. 
An alternative justification of a time-dependent-only $\lambda$
is based on re-parametrization invariance,
which has been fruitful in justifying the known classical 
long-range forces and some of the key properties of physical systems
\cite{RepInv-and-KeyProps-of-Phys-Sys'21}.
Either way, one arrives at the above arguments and concludes
that the Einstein Cosmological Constant $\Lambda_E$
is a manifestation of the choice of parametrization where the 
age of the Universe is a natural parameter that measures
how much stuff has become causally connected within the
observed Universe. Thus, upon a proper choice of $\lambda$ and
metric $g_{\mu\nu}$, the $\Lambda_E$ term disappears, and 
one has only \eqref{eq: No Lambda} without the observer-related 
cosmological constant $\Lambda_E$ given by \eqref{LambdaE}.

\section{The Missing Mass Problem}

At astronomical and cosmological scales, it has been observed that
motions in galaxies and their clusters exhibit behavior
that cannot be explained by the observed known matter and its laws
of motion. The idea of an invisible dark matter has been proposed
to explain observations (see the next section for more
details). However, its absence from laboratory tests for the past
40 years has become a puzzle. That is, the absence of any laboratory
detection of new dark matter particles raises questions as to the
validity of the idea of the existence of dark matter. On the other
hand, these observations could be addressed by an alternative
to the dark matter hypothesis known as Modified Newtonian Dynamics
(MOND) \cite{MOND-as-DM-alternative'83}. 
It has been gaining supporters due to its successes in fitting
observational data about galaxies \cite{MOND-laws'14}. 
Furthermore, it has been shown that MOND 
could be viewed as a particular manifestation of
the SIV paradigm \cite{MOND-from-SIV'23}. 
Thus, the presence of the scale factor $\lambda$
within the SIV paradigm can be utilized to address this 
missing matter observational puzzle by its connection to MOND.

\subsection{The Dark Matter Option}

There is unambiguous observational evidence for deviation away from
the Keplerian fall-off $v^{2}\sim M/r$ when observing the motion
of stars in the outer layers of galaxies.
The observations point to flat rotational curves $v\sim const$
as one moves further away from the central bulge of the visible part of a galaxy. 
This observation has been addressed as a continuation
of the matter paradigm by assuming the presence of extra non-luminous
matter named dark matter, which forms halos around galaxies and extends
far beyond the luminous part of the host galaxy \cite{DM-History-Bertone'18}. 
This is a natural initial guess as to 
what may be causing the deviation from the Keplerian
fall-off and into the presence of flat rotational curves. 
There is a large variety of proposed dark matter candidates that are yet
to be observed if the idea is correct. In addition to the fact that such
dark matter is a no-show in labs, there are many additional issues
with the various proposed dark matter solutions.

\subsection{The MOND Option}

The Modified Newtonian Dynamics (MOND) resolution to the observed
flat rotational curves does not consider a new dark matter component(s)
but assumes that the dynamics are changed once the Newtonian acceleration
falls below a certain cut-off value $a_{0}$. 
Since the FLRW scale factor at the current epoch is often chosen to be 1,
this ambiguity in notation with the MOND acceleration $a_{0}$
should presumably be absent.

For the very low acceleration $g\ll\,a_0$, the initial MOND suggest that
$g\sim \sqrt{g_{N}a_0}$, where $g=v^{2}/r$ is the observed
acceleration, while $g_{N}=GM/r^{2}$ is the Newtonian gravitational
acceleration \cite{MOND-as-DM-alternative'83}. 
For systems with acceleration $g\gg a_0$, 
the dynamics are reduced to the standard dynamics $g=g_{N}$
(limit of $a_0\rightarrow0$), while, for accelerations $g\ll a_0$,
the system is in the Deep-MOND (DMOND) regime, where it also should exhibit
scale-invariant space--time dynamics $(t,r)\rightarrow\lambda\times(t,r)$
\cite{MOND_and_ScaleInv'09,MOND-laws'14}.
Furthermore, in the scale-invariant DMOND regime, $g\sim\sqrt{g_{N}a_0}$,
where the limit could be viewed as $a_0\rightarrow\infty$ along
with $G\rightarrow0$ but the product $Ga_0\rightarrow\mathcal{A}_{0}$
stays constant. In this DMOND scale-invariant regime, $\lambda$
is an overall constant scale factor, where $(t,r)\rightarrow\lambda\times(t,r)$
could be viewed as rescaling the coordinate functions but keeping
the units the same; thus, $G$ and $a_0$ are constants that are
invariant upon this rescaling within the DMOND regime. An alternative
viewpoint is to consider a change of units that removes the $\lambda$
scaling of the coordinates while inducing change in the values of
the dimension-full constants $q$, say with units $[q]=[l]^{a}[t]^{b}[m]^{c}$,
then $q\rightarrow\lambda^{-(a+b)}q$ \cite{MOND-laws'14}. 
From this viewpoint, $a_0\rightarrow\lambda a_0$ and $G\rightarrow G/\lambda$ and 
the limits are obtained as $\lambda\rightarrow0$ or $\infty$. 
Notice that mass-related quantities are not affected within MOND. 
The value of the MOND acceleration $a_0$ is expected
to satisfy $a_0\approx cH_{0}/2\pi$ \cite{MOND-laws'14}.

\subsection{Deriving MOND-like Acceleration Within SIV}

In the SIV paradigm, the scale invariance is the primary idea, and the
form of the scale factor $\lambda$ can be deduced, as discussed earlier. 
Furthermore, the equations of motion given by the equations
of the geodesics within GR are generalized via Dirac co-calculus
to include an extra velocity-dependent term 
as part of the scale-covariant Newtonian equation of motion
\citep{1978A&A....65..337M,WIG'78,ScaleInv_and_AstroMotion'79,Maeder'17c,SIV-PartI'23}:
\begin{equation}
\frac{d^{2}\overrightarrow{r}}{dt^{2}}
=-\frac{G_{t}M(t)}{r^{2}}\frac{\overrightarrow{r}}{r}+\kappa(t)\frac{d\overrightarrow{r}}{dt},
\label{eq: SIV dynamics}
\end{equation}

Here, $\kappa(t)=-\dot{\lambda}/\lambda$ and $G_{t}$ is Newton's
Gravitational Constant in the system of units related to the choice
of time parametrization $t$ within the SIV, where $G_{t}$ is viewed as a true
constant but the mass is expected to have relevant time dependence.
For the above equation to exhibit scale invariance, one has
to assume that $G_{t}M(t)\propto\lambda$, and if $G_{t}$ is kept
constant, then $M(t)=M_{0}\lambda(t)$. 
The variation of mass is demanded by the conservation law 
associated with the scale-invariant equation 
\cite{Maeder'17a}.
In the subsequent considerations, we will drop the time label for 
$G_{t}$ and $M(t)$ to simplify the notation.

To arrive at an expression for an MOND-like 
acceleration $a_0$ within the SIV, one considers the ratio of the 
Newtonian acceleration $g_{N}=GM/r^{2}$ to the additional dynamic acceleration $\kappa(t)v$
(magnitudes):
\begin{equation}
x=\frac{\kappa vr^{2}}{GM}.\label{eq: x}
\end{equation}

Next, we will use the relation given by the instantaneous radial acceleration
$v^{2}/r=GM/r^{2}$ to eliminate the speed $v$. Then, by using
$g_{N}=GM/r^{2}$, we arrive at
\[
x=\frac{\kappa vr^{2}}{GM}=\kappa\sqrt{\frac{r^{3}}{GM}}=\kappa\sqrt{\frac{r}{g_{N}}}.
\]

The quantity $x$ has been discussed previously 
\cite{Maeder'17c,SIVandDS'20}, 
{and} is 
finally utilized in connecting the SIV to MOND in \cite{MOND-from-SIV'23}.
Here, we are re-deriving the relevant expressions with a focus on keeping 
the time component $\kappa=\kappa_0=-\dot{\lambda}/\lambda$ 
of the connexion vector within the WIG explicit.
Thus, when the dynamic acceleration dominates over the Newtonian ($x\gg1$),
one has
\[
g=g_{N}+xg_{N}\approx xg_{N}=\kappa\sqrt{rg_{N}}.
\]

Therefore, we have arrived at the DMOND-type relation $g\sim\sqrt{a_0g_{N}}$,
from which we can deduce an expression for $a_0$ within the SIV:
\[
a_0\approx\kappa^{2}r.
\]

The upper bound on $a_0$ corresponds to utilizing the Hubble horizon
$r\rightarrow r_{H}=c/H_{0}$. Therefore, 
\begin{equation}
a_0\approx\kappa^{2}r_{H}=\kappa^{2}c/H_{0}.\label{eq: a_0}
\end{equation}

In SIV units ($1=\lambda_{0}=t_{0}=c$), one has $\text{\ensuremath{\kappa}}=1/t$,
and, for the matter-dominated epoch, using the scale factor 
$a(t)=\left((t^{3}-\Omega_\text{m})/(1-\Omega_\text{m})\right)^{2/3}$ \cite{SIV-Cosmology-Jesus'19},
one obtains $H=2t^{2}/(t^{3}-\Omega_\text{m})$; therefore, 
$a_0\approx\kappa^{2}c/H_{0}=(1-\Omega_\text{m})/2=\Omega_{\lambda}/2$,
where, within the SIV, $\Omega_{\lambda}=2/(H\,t)$---assuming a flat Universe ($\Omega_\text{k}=0$).
Therefore, $\Omega_{\lambda}+\Omega_\text{m}=1$ is trivially true within the SIV.
SIV models with non-zero curvature $k$ are also possible \cite{Maeder'17a}.
Thus, a non-zero $\kappa$ term in \eqref{eq: SIV dynamics} gives rise to 
non-zero DMOND-like acceleration within the SIV:
\begin{equation}
a_0\approx\Omega_{\lambda}/2.
\end{equation}

Next, we look at the DMOND-like acceleration $a_0$, as expressed
in the usual SI units where the age of the Universe is $\tau_{0}=13.8$ billion years, the
Hubble constant is $H_{0}=68$ km/s/Mp, and the speed of light is $c=3\times10^8$ m/s.
Using $\kappa\approx1/\tau_{0}$, one can immediately estimate the
value to be $a_0\approx c/\tau_{0}$ since $H_{0}\tau_{0}\approx1$.
However, to be more precise, one has to take into account that 
$\kappa=-\dot{\lambda}/\lambda$ and therefore $\kappa(\tau)=(dt/d\tau)\kappa(t)$. 
To find $dt/d\tau$, we consider
\[
\frac{t-t_\text{in}}{t_{0}-t_\text{in}}=\frac{\tau-\tau_\text{in}}{\tau_{0}-\tau_\text{in}}
\]
for $\tau_\text{in}=0$ and $t_\text{in}=\Omega_\text{m}^{1/3}$
$ \Rightarrow t=\Omega_\text{m}^{1/3}+\frac{\tau}{\tau_{0}}(1-\Omega_\text{m}^{1/3})$, 
\[
\frac{dt}{d\tau}=\frac{t_{0}-t_\text{in}}{\tau_{0}}=\frac{(1-\Omega_\text{m}^{1/3})}{\tau_{0}}.
\]

Thus, there is a correction factor to our use of $\kappa\approx1/\tau_{0}$ above;
that is, $\kappa(\tau_{0})=(1-\Omega_\text{m}^{1/3})/\tau_{0}$ results in
\[
a_0\approx(1-\Omega_\text{m}^{1/3})^{2}c/\tau_{0}=(1-\Omega_\text{m}^{1/3})^{2}cH_{0}/\xi,
\]
where $H_0\tau_0=\xi\approx1$.
For $\Omega_\text{m}=5\%$, this gives $a_0\approx 2.75\times 10^{-10}$ m/s$^2$.
Here, the estimate is based on the $\Lambda$CDM model; its fit to observational data 
results in $\Omega_\text{m}=5\%$ for baryonic matter. 
Within the SIV, we do not have such parameter determination yet;
however, a value of $\Omega_\text{m}$ could be estimated based on the self-consistency requirement  
about the age of the Universe and the value of the Hubble constant. That is, assuming
$H_0\tau_0=\xi\approx1$, one obtains $2(1-\Omega_\text{m}^{1/3})/(1-\Omega_\text{m})\approx1$,
which gives $\Omega_\text{m}\approx23.6\%$ within the SIV. Thus, $a_0\approx 10^{-10}$ m/s$^2$.
Therefore, the two estimates above for $a_0$ 
result in about the same order of magnitude values
$a_0\approx10^{-10}$ m/s$^2$.
Notice that this value for $\Omega_\text{m}=23.6\%$ 
is closer to the total matter content within the $\Lambda$CDM model.

The SIV paradigm suggests that the MOND-like 
acceleration $a_0$ may be epoch dependent 
and could have different values depending on the redshift of the system under observation.
This could be used to differentiate and test the SIV and MOND paradigms
\cite{MOND-a(z)-dependence-Del_Popolo'24}.

\section{Conclusion}
To summarize, we have shown that the Einstein Cosmological Constant $\Lambda_E$ 
is a true constant within the framework of scale-invariant cosmology. 
In particular, within the SIV paradigm, one can derive an expression \eqref{eq: lambda} 
for the scale factor $\lambda(t)$ based on \eqref{eq: SIV1}, 
which results in the condition \eqref{eq: Lambda SIV},
while the metric tensor satisfies the usual Einstein equation without
cosmological constant \eqref{eq: No Lambda}. 
That is, without a cosmological constant term of the form $\Lambda_E g_{\mu\nu}$, 
but with the potential introduction of time-dependent dark energy term $\tilde{T}_{\mu\nu}$, 
that could be significant in the early universe and then diminish later.
The contemporary view on the resolution of the Hubble tension is the possibility for early dark energy \cite{EDEandHT'23},
which could be supplied by $\tilde{T}_{\mu\nu}$.
This explicit new model for early dark energy has the form $\tilde{T}_{\mu\nu}\sim\kappa\,H$;
thus, it could be used to test the SIV theory and its impact on the Hubble tension.
However, one first will have to determine the model parameter $\Omega_m$ and the validity of the SIV paradigm.
This could be performed by determination of the relevant SIV $\Omega_m$ 
from the cosmological parameters such as deceleration, jerk, and snap,  
which were recently constrained using a model-independent kinematic cosmographic study 
utilizing three different data sets and their combinations \cite{2024A&A...690A..40L, 2024arXiv240412068C}.
Our approach to $\Lambda_E$ avoids the puzzling conclusion that 
we may be inside of a black hole if $\Lambda_E$ is associated with 
the zero-point energy of the vacuum---labeled as dark energy that does not gravitate.
Furthermore, we avoid the puzzling observation that there is a disagreement of 123 orders of magnitude with the QFT estimates of the zero-point vacuum energy since it does not contribute to $T_{\mu\nu}$.
In our approach, the presence of a non-zero $\Lambda_E$ within $\Lambda$CDM is due to 
the choice of time parameterization. That is, as there is extra kinetic energy in the case of spatial motion,
there is also extra energy due to differences in the time parametrization 
since this is a relative temporal motion.

Next, we have shown how to derive an expression for the MOND-like 
acceleration $a_0$ \eqref{eq: a_0}
that controls the transition to DMOND where scale invariance is expected.
In this respect, we may have explained the dark matter problem via 
an MOND-like paradigm.
In our case, however, $a_0$ is a consequence of the scale-invariant equations of motion 
\eqref{eq: SIV dynamics}, while the usual MOND limits are controlled by the parameter $x$ \eqref{eq: x}.
Thus, there is a small tangency 
of the SIV and MOND over a  
limited interval of low gravities and timescales!

Our estimates of the fundamental MOND acceleration $a_0\approx 10^{-10}$ m/s$^2$ and the 
Einstein Cosmological Constant $\Lambda_{E}\approx 10^{-52}\;\text{m}^{-2}$ show the correct order of 
magnitude for these two important constants. Both values are controlled by the age of the Universe,
while $a_0$ is also related to the matter content of the Universe.

\vspace{6pt}

\authorcontributions{Writing---original draft V.G.G.;
conceptualization---both authors;
formal analysis---both authors;
investigation---both authors;
methodology---both authors;
validation---both authors;
writing---review and editing---both authors.
Both co-authors have been actively involved in the writing of the paper and its draft versions. 
All authors have read and agreed to the published version of the~manuscript.}

\funding{This research received no external~funding.}

\dataavailability{No new data were created or analyzed in this study.} 



\acknowledgments{
This research did not receive any specific grant from funding agencies in the public, commercial, or not-for-profit sectors.
A.M. expresses his gratitude to his wife for her patience and support. 
V.G.G. is extremely grateful to his wife and daughters for their understanding and family support  during the various stages of the research presented.}

\conflictsofinterest{The authors declare no conflicts of~interest.} 






\begin{adjustwidth}{-\extralength}{0cm}

\printendnotes[custom]
\reftitle{References}


\PublishersNote{}
\end{adjustwidth}
\end{document}